\documentclass[12pt]{article}

\usepackage{amsmath}

\usepackage{amsfonts}

\usepackage{amssymb}

\usepackage[latin1]{inputenc}

\usepackage{graphics}

\usepackage{graphicx}

\usepackage{epsfig}

\usepackage{amsthm}

\usepackage{lineno}

\usepackage{mathrsfs}

\usepackage{latexsym}

\newcommand{\h}{\hspace{.5cm}}
\newenvironment{destaque}{\begin{quotation}\small\em}{\end{quotation}}

\date{}

\begin{document}

\title{Ground State of the Hydrogen Atom via Dirac Equation in a Minimal Length Scenario}
\author{{\bf T. L. Antonacci Oakes, R. O. Francisco, J. C. Fabris and}\\
 {\bf J. A. Nogueira}\footnote{e-mail: jose.nogueira@ufes.br}\\
{\small \it Departamento de F\'{\i}sica}\\
{\small \it Universidade Federal do Esp\'{\i}rito Santo}\\
{\small \it 29.075-910 - Vit\'oria-ES - Brasil}}

\maketitle

\begin{abstract}
\begin{destaque}
\h In this work we calculate the correction to the ground state energy of the hydrogen atom due to contributions arising from the presence of a minimal length. The minimal length scenario is introduced by means of modifying the Dirac equation through a deformed Heisenberg algebra (kempf algebra). With the introduction of the Coulomb potential in the new Dirac energy operator, we calculate the energy shift of the ground state of the hydrogen atom in first order of the parameter related to the minimal length via perturbation theory.
\end{destaque}
\end{abstract}



\section{Introduction}
\label{introd}

\h It is a remarkable fact that all approaches to quantum gravity seem to coincide in one prediction: the existence of a \emph{minimal length}. However the idea of the existence of a minimal length is very before any attempt to quantizing gravity. Due to the divergences arising from the advent of Quantum Field Theory, in the 1930s, W. Heisenberg concluded that a fundamental length should exist which would be as a natural cut-off for divergent integrals \cite{kragh1,Heisenberg}. Heisenberg tried, without success, to make sense of a fundamental length by allowing that the components of the position operator do not commute. But it was not until 1947, that H. S. Snyder proposed a Lorentz-covariant algebra of the position and momentum operators in which the components of the position operator do not commute. The Snyder's proposal leads to a non-continuous space-time, and, in this way, a minimal length is introduced in theory \cite{Snyder}. In the 1994, S. Majid and H. Ruegg proposed a modification for the commutation relations of space-time coordinates which became known as $\kappa$-Poincar\'e Algebra \cite{Majid}. In the same year, A. Kempf, G. Mangano and R. B. Mann initiated the development of the mathematical basis of quantum mechanics in a minimal length scenario \cite{Kempf:1994su}. As far as is known, M. Bronstein was the first to realize that the quantization of gravity leads to a limit for the precision of a measurement, and consequently the existence of a minimal length \cite{Bronstein}. But it was only in 1964 that C. A. Mead recognized the relevant role that gravity plays in trying to probe a smaller and smaller region of the space-time \cite{Mead1}.  Over the years, many works have been published about minimal length in different contexts. For more about ideas of the existence of a minimal length and its implementation, the interested reader is referred to references \cite{Mead2,chang,Sprenger,Hossenfelder0}.

The inclusion of a minimal length in the theoretical framework has been accomplished through different ways \cite{Hossenfelder1}: generalization of the uncertainty principle (GUP), deformation of the special relativity (DSR) and modification of the dispersion relation (MDR).

In quantum theory, the existence of a minimal length can be described as a non-zero minimal uncertainty $\Delta x^{min}$ in the measurement of position, which leads to a generalization of the Heisenberg's uncertainty principle. Several generalizations of the uncertainty principle have been considered in the literature. Kempf et al. \cite{Kempf:1994su,Kempf2:1997} proposed a three-dimensional generalized uncertainty relation, which implements the appearance of a non-zero minimal uncertainty in position, of the form\footnote{We use boldface to a vector operator for a sake of simplicity.}
\begin{equation}
	\label{rc1kempf}
	[\hat{X}_i,\hat{P}_j] = i\hbar \left[  \left(1 + \beta\hat{\bf P}^2 \right) \delta_{ij} + \beta^{\prime}\hat{P}_{i}\hat{P}_{j} \right],
\end{equation}
where $\beta$ and $\beta^{\prime}$ are parameters related to the minimal length and $\hat{\bf P}^2 = \sum^{3}_{i=1} \hat{P}^{2}_{i}$.
If the components of the momentum operator are assumed to commute with each other,
\begin{equation}
	\label{rcM}
	[\hat{P}_i,\hat{P}_j] = 0,
\end{equation}
then the commutation relations among the components of the position operator are determined by the Jacobi identity as\footnote{There is a summation over dummy indices.}
\begin{equation}
	\label{rc2kempf}
	[\hat{X}_i,\hat{X}_j] = -i\hbar \left[ 2\beta - \beta^{\prime} + \left(2\beta + \beta^{\prime} \right)\beta\hat{\bf P}^2 \right] \epsilon_{ijk}\hat{L}_{k},
\end{equation}
where
\begin{equation}
	\label{rc3kempf}
	\hat{L}_{i} = \frac{1}{\left(1+ \beta\hat{\bf P}^2 \right)} \epsilon_{ijk}\hat{X}_{j}\hat{P}_{k},
\end{equation}
are the components of the orbital angular momentum operator, satisfying the usual commutation relations $[\hat{L}_{i},\hat{X}_{j}] = i \hbar \epsilon_{ijk}\hat{X}_{k}$ and $[\hat{L}_{i},\hat{P}_{j}] = i \hbar \epsilon_{ijk}\hat{P}_{k}$. This algebra gives rise to (isotropic) non-zero minimal uncertainties in the position coordinates $\Delta X^{min}_{i} = \hbar \sqrt{3 \beta + \beta^{\prime}}$, as it was expected. Here it would be interesting to mention that non-zero commutation relations for the components of the position operator can be obtained from a classical model for particles with spin without introducing a minimal length \cite{Suttorp,Kerf}. 

A big challenge has been the search for experimental constraints to obtain an upper bound for the minimal length value. Such experimental constraints are particularly relevant to models of large dimensions which possess a much lower effective Planck scale than 4-dim value \cite{arkani,appelquist,Hossenfelder2}. The corrections to the energy spectrum of the hydrogen atom due to the presence of a minimal length have been calculated by many authors \cite{Hossenfelder2,brau,yao,benczik,nouicer,bouaziz,samar}. The accuracy concerning the experimental measurement of the frequency of the radiation emitted during the transition 1S-2S was used for the first time by Brau \cite{brau} to estimate a maximum value to minimal length of order $10^{-17}m$. Considering the Schr\"odinger equation for the hydrogen atom, Brau has found that the contribution to the ground state energy of the hydrogen atom, which arises due to the presence of the minimal length, is of order ${\cal O}(\alpha^{4})$, where $\alpha$ is the fine structure constant.

It would be interesting to study the effects of the relativistic contributions due to the presence of a minimal length on the hydrogen atom. This means we will have to consider the Dirac equation for the hydrogen atom in a minimal length scenario. Thus we use the Kempf algebra to introduce the Dirac equation in a minimal length scenario in order to calculate the energy of the ground state of the hydrogen atom. 

The article is organized in the following way. In section \ref{Deoml}  we derive the new Dirac energy operator in a minimal length scenario taking into account the Kempf algebra.  In section \ref{Fsha} we calculate the energy of the ground state of the hydrogen atom in a minimal length scenario and roughly estimate an upper bound for the value of the minimal length. We present our conclusions in section \ref{Concl}. In Appendix \ref{ree} we show how we have calculated the relativistic energy of the electron and in Appendix \ref{B} we outline how we have calculated the corrections in the first order of perturbations.


\section{Dirac Energy Operator in a Minimal Length Scenario}
\label{Deoml}

\h In this section we derive the Dirac energy operator taking into account the Kempf algebra. A quick glance at Eq.(\ref{rc2kempf}) shows that in the special case $\beta^{\prime} = 2\beta$ the components of the position operator commute to first order of the minimal length parameter. Therefore we regard the case $\beta^{\prime} = 2 \beta$. Then the Eqs (\ref{rc1kempf}), (\ref{rcM}) and (\ref{rc2kempf}) to first-order of $\beta$ become
\begin{equation}
\label{rc1}
	[\hat{X}_i,\hat{P}_j] = i\hbar \left[  \left(1 + \beta\hat{\textbf{P}}^2 \right) \delta_{ij} + 2\beta\hat{P}_{i}\hat{P}_{j} \right],
\end{equation}
\begin{equation}
\label{rc2}
	[\hat{P}_i,\hat{P}_j] = 0,
\end{equation}
\begin{equation}
\label{rc3}
	[\hat{X}_i,\hat{X}_j] = 0.
\end{equation}
The commutation relations above lead to minimum $\Delta X^{min}_{i} = \hbar \sqrt{5 \beta}$. It is not difficult to verify that the following representation (which we call ``position'' representation) fulfil the relations above to first order in $\beta$,
\begin{equation}
\label{rx1}
	\hat{\bf X} =\hat{\bf x},
\end{equation}
\begin{equation}
\label{rp1}
	\hat{\bf P} \equiv \left( 1 - \beta \hbar^{2} \nabla^{2} \right)(-i\hbar \vec{\nabla}).
\end{equation}

Note that $\hat{\bf x}$ and $\hat{\bf p} \equiv -i\hbar \vec{\nabla}$ are position and momentum operators in ordinary quantum mechanics\footnote{We use ``ordinary quantum mechanics'' in opposition to quantum mechanics in a minimal length scenario.}, that is, $\hat{x}_{i}$ and $\hat{p}_{i}$ satisfy
\begin{equation}
	[\hat{x}_i,\hat{x}_j] = 0,
\end{equation}\begin{equation}
	[\hat{p}_i,\hat{p}_j] = 0,
\end{equation}
\begin{equation}
\label{rco1}
	[\hat{x}_i,\hat{p}_j] = i\hbar \delta_{ij}.
\end{equation}

The Dirac equation in the ordinary quantum mechanics is 
\begin{equation}
\label{edirace}
i\hbar \frac{\partial | \psi \rangle}{\partial t} = \left[ c \left(\vec{\alpha} \cdot \hat{\bf p} \right) + \hat{\beta}mc^{2} \right]| \psi \rangle ,
\end{equation}
where
\begin{equation}
\hat{\beta} = 
\begin{pmatrix}
1 & 0 \cr
0 & -1 \cr
\end{pmatrix},
\end{equation}
\begin{equation}
\vec{\alpha} = 
\begin{pmatrix}
0 & \vec{\sigma} \cr
\vec{\sigma} & 0 \cr
\end{pmatrix},
\end{equation}
and $\vec{\sigma}$ are the Pauli matrices\footnote{$\vec{\alpha}$ and $\hat{\beta}$ must be not confused with the fine structure constant $\alpha$ and the minimal length parameter $\beta$.}. So the energy operator is given by
\begin{equation}
\label{edirace}
\hat{E} = c \left(\vec{\alpha} \cdot \hat{\bf p} \right) + \hat{\beta}mc^{2}.
\end{equation}

In order to introduce a minimal length in theory we replace $\hat{p_i}$ by $\hat{P_i}$. Thus,
\begin{equation}
\label{rE1}
\hat{E}_{ML} = c \left(\vec{\alpha} \cdot \hat{\bf P} \right) + \hat{\beta}mc^{2},
\end{equation}
where $\hat{E}_{ML}$ is the new energy operator.

Now, using the Eq.(\ref{rp1}) we can rewrite $\hat{E}_{ML}$ as
\begin{equation}
\label{rE3}
\hat{E}_{ML}= c\left(\vec{\alpha} \cdot \hat{\bf p} \right) + \hat{\beta}mc^{2}  +	\beta c\left( \vec{\alpha} \cdot \hat{\bf p} \right)^{3},
\end{equation}
where we have used the relation $\hat{\bf p}^{2} = \left(\vec{\alpha} \cdot \hat{\bf p} \right)^{2}$.

We must note that the new momentum operator $\hat{\bf P}$ no longer coincides with the generator of space translation $-i\vec{\nabla}$, as well as the new energy operator $\hat{E}_{ML}$ no longer coincides with the generator of time translation $i\frac{\partial}{\partial t}$ \cite{Kempf1}. They are related by \cite{Hossenfelder2}
\begin{equation}
\label{Et}
\hat{E}_{ML} \equiv i \hbar \frac{\partial}{\partial t} \left( 1 + \beta \hbar^{2} \frac{\partial^{2}}{\partial t^{2}} \right),
\end{equation}
or 
\begin{equation}
\label{tE}
i \hbar \frac{\partial}{\partial t} \equiv \hat{E}_{ML} \left( 1 - \frac{\beta}{c^2} \hat{E}^{2}_{ML} \right).
\end{equation}

Using the relation (\ref{Et}) between $\hat{E}_{ML}$ and $i \hbar \frac{\partial}{\partial t}$ operators into Eq.(\ref{rE3}) we get
\begin{equation}
\label{Dirac1}
\left( i \hbar \frac{\partial}{\partial t} + i \beta \hbar^{3} \frac{\partial^{3}}{\partial t^{3}} \right) | \psi_{ML} \rangle =
\left[ -i \hbar c\left(\vec{\alpha} \cdot \vec{\nabla} \right) + \hat{\beta}mc^{2}  +	i \beta \hbar^{3} c\left( \vec{\alpha} \cdot \vec{\nabla} \right)^{3} \right]
 | \psi_{ML} \rangle ,
\end{equation}
which treats space and time in a manifestly symmetric fashion. However, we can write the Dirac equation for a first order time derivative by using the Eq.(\ref{rE1}) and the energy mass shell condition, $\hat{E}^{2}_{ML} = c^{2}\hat{\bf P}^{2} + m^{2}c^{4}$, into Eq.(\ref{tE}),
\begin{equation}
\label{Dirac2}
i \hbar \frac{\partial}{\partial t} | \psi_{ML} \rangle = \left[ c \left(\vec{\alpha} \cdot \hat{\bf P} \right) + \hat{\beta}mc^{2} \right] 
\left[ 1 - \beta \left( c^{2}\hat{\bf P}^{2} + m^{2}c^{4} \right) \right] | \psi_{ML} \rangle ,
\end{equation}
where $\hat{\bf P}$ is given by Eq.(\ref{rp1}).

It is clear that in cases which we are interested in the energy spectrum it is more convenient to employ the Eq.(\ref{rE3}).


\section{Energy of the hydrogen atom ground state}
\label{Fsha}

\h With the intention of finding out the energy of the ground state of the hydrogen atom in a minimal length scenario we introduce the electrostatic central potential of the proton in Eq.(\ref{rE3}). Because of Eq.(\ref{rx1}) the central potential is not modified in order which we are considering, i. e., for $\cal O(\beta)$. So, the energy equation is\footnote{Note that $x_{i}$ is not eigenvalue of the $\hat{X}_{i}$ operator. In fact, the existence of the minimal length implies that $\hat{X}_{i}$ operator can not have any eigenstate which is a physical sate, that is, any eigenfunction within the Hilbert space \cite{Kempf:1994su,dorsch}. Nevertheless, the ``position'' representation is particularly useful when the shifts in the energy can be calculate via perturbation theory in X-space \cite{chang}.}
\begin{equation}
\label{rEcp}
E_{ML}| \psi_{ML} \rangle =
 \left[ c \left(\vec{\alpha} \cdot \hat{\bf p} \right) + \hat{\beta}mc^{2} -\frac{\hbar c \alpha}{r} + 
 \beta c \left( \vec{\alpha} \cdot \hat{\bf p} \right)^{3} \right]| \psi_{ML} \rangle
\end{equation}

If we assume the mass scale of the minimal length $M_{ML}$ to be big so that the electron mass is much smaller than it ($\beta = \frac{c^{2}}{M^{2}_{ML}c^{4}}$, therefore $\beta m^{2}c^{2} = \frac{m^{2}}{M^{2}_{ML}} \ll 1$), then we can consider the fourth term as a perturbation. The evaluation of the energy to first order in $\beta$ leads to
\begin{equation}
E_{ML} = E + \beta m^{2}c^{2}E^{1},
\end{equation}
where $E$ is the energy of the $ | \psi \rangle $ state of the hydrogen atom obtained from the ordinary Dirac equation and $E^{1}$ is given by
\begin{equation}
E^{1} = \langle \psi | \frac{1}{m^{2}c} \left( \vec{\alpha} \cdot \hat{\bf p} \right)^{3} | \psi \rangle.
\end{equation}

Although the explicit calculation of the above expression is very laborious, as can be seen in Appendix B, we carry it out to find the ground state energy of the hydrogen atom. We obtain the following result (see Appendix B1)
\begin{equation}
\label{er}
E_{0}^{ML} = mc^{2}\epsilon + \beta m^{3}c^{4} \frac{\left(1 - \epsilon^{2} \right)^{2}}{\epsilon\left(2\epsilon - 1 \right)},
\end{equation}
where  $\epsilon = \sqrt{1 - \alpha^2} $.

It is interesting to expand $E_{0}^{ML}$ in power of the fine structure constant. After we perform some simple calculations and subtract the rest energy of the electron,
\begin{equation}
\Delta E_{0}^{ML} = E_{0}^{ML} - mc^{2},
\end{equation}
we get
\begin{equation}
\label{edml}
\Delta E_{0}^{ML} \approx - mc^{2} \left( \frac{\alpha^{2}}{2} + \frac{\alpha^{4}}{8} \right) +  
\beta m^{3}c^{4} \alpha^{4}.
\end{equation}

This result shows that the correction to the energy of the ground state of the hydrogen atom is always positive and it has the same order in fine structure constant of the result obtained by Brau.

We can make a estimation of the minimal length value comparing our theoretical result with the experimental data. As far as we know, the best accuracy concerning the measurement of the 1S-2S energy splitting in the hydrogen atom has been obtained by C. G. Parthey et al \cite{parthey}. They have gotten an accuracy of about $4,2 \times 10^{-14}$ eV (2,466,061,413,187,035(10)Hz, an accuracy of 4 parts in $10^{15}$). The calculation of the energy of the 2S state is a very boring job. However, in order to make a rough estimation, we do not need to find the exact energy of the 2S state, since the contribution of the lowest order to the correction of the energy of the 2S state must be of ${\cal O}(\alpha^{4})$, because the 1S and 2S ordinary states have the same symmetry. Therefore the energy difference between the 1S and 2S states for minimal length correction must lead to a result of ${\cal O}(\alpha^{4})$, since the 1S and 2S levels differ only in the $n$ quantum number. If we attribute this error entirely to the minimal length corrections and assume that the effects of the minimal length can not yet be seen experimentally, from (\ref{edml}), we find 
\begin{equation}
\Delta X_{i}^{min} \leq 10^{-17} m.
\end{equation}
As it was expected the result is identical to one obtained by Brau \cite{brau}.


\section{Conclusion}
\label{Concl}

\h The aim of this work was to calculate, in a relativistic approach, the energy of the ground state of the hydrogen atom in a minimal length scenario. The minimal length has been introduced in the theory through the generalization of the Heisenberg's algebra chosen by Kempf and in the special case $\beta^{\prime} = 2\beta$, see Eqs. (\ref{rc1}), (\ref{rc2}) and (\ref{rc3}). In order to avoid the problem of substituting $\hat{X}_{i}$ for derivatives of $\hat{p}_{i}$ in the Coulomb potential ($\frac{1}{r}$) we have used the ``position'' representation given by equations (\ref{rx1}) and (\ref{rp1}). With the introduction of the Coulomb potential in the new Dirac energy operator obtained by replacement of $\hat{p_i}$ by $\hat{P_i}$ in the ordinary Dirac Hamiltonian, we have found a expression for the energy shift of the hydrogen atom via perturbation theory, since we have assumed that the electron mass is much smaller than the mass scale of the minimal length (see Eq.(\ref{3fi3}) in Appendix B). Moreover, we have explicitly calculated the ground state energy of the hydrogen atom and found that it is of ${\cal O}(\alpha^{4})$. Comparing our result with experimental data we can roughly estimate the upper bound for the minimal length value of the order of $10^{-17}m$, as we expected it is exactly equal to the Brau's result.



\section*{Acknowledgements}

\h We would like to thank CAPES, CNPq and FAPES (Brazil) for financial support.
\\
\\
\\

\section*{Appendix}
\numberwithin{equation}{section}

\appendix


\section{Relativistic energy of the electron}
\label{ree}

\h From Eq.(\ref{rE3}) we have a linear homogeneous system of equations for $\phi$ and $\chi$,
\begin{equation}
\begin{pmatrix}
- E_{ML} + mc^{2} & 
c \left( \vec{\sigma} \cdot \hat{\bf p} \right) +  \beta c \left( \vec{\sigma} \cdot \hat{\bf p} \right)^{3} \cr
c \left( \vec{\sigma} \cdot \hat{\bf p} \right) +  \beta c \left( \vec{\sigma} \cdot \hat{\bf p} \right)^{3} & 
- E_{ML} - mc^{2} \cr
\end{pmatrix}
\begin{pmatrix}
\phi \cr
\chi \cr
\end{pmatrix}
= 0,
\end{equation}
which has non-trivial solution only for
\begin{equation}
\begin{vmatrix}
- \left( E_{ML} - mc^{2} \right) & 
c \left( \vec{\sigma} \cdot \hat{\bf p} \right)  \left( 1 + \beta  p^{2} \right) \cr
c \left( \vec{\sigma} \cdot \hat{\bf p} \right)  \left( 1 + \beta  p^{2} \right) & 
- \left( E_{ML} + mc^{2} \right) \cr
\end{vmatrix}
= 0.
\end{equation}

Thus, after throwing away terms of order $\beta^{2}$, the relativistic energy of the free electron in the regarded scenario of minimal length can be obtained from 
\begin{equation}
\label{EA1}
E^{2}_{ML} = p^{2}c^{2} + m^{2}c^{4} + 2\beta c^{2} p^{4},
\end{equation}
as it was expected from $E^{2}_{ML} = c^{2}P^{2} + m^{2}c^{4}$.

\section{Corrections in the first order of perturbation}
\label{B}
\h In order to obtain
\begin{equation}
\langle \psi | \left( \vec{\alpha} \cdot \hat{\bf p} \right)^{3} | \psi \rangle =
\int  \left(\phi_{1}^{\dag}, \phi_{2}^{\dag} \right)
\begin{pmatrix}
0 & \left( \vec{\sigma} \cdot \hat{\bf p} \right)^{3} \cr
\left( \vec{\sigma} \cdot \hat{\bf p} \right)^{3} & 0 \cr
\end{pmatrix}
\begin{pmatrix}
\phi_{1} \cr
\phi_{2} \cr
\end{pmatrix}
d^{3}\vec{x},
\end{equation}
where $\phi$ and $\chi$ are two-component eignspinors of the state,
\begin{equation}
\langle \vec{x} | \psi \rangle =
\begin{pmatrix}
\phi_{1} \cr
\phi_{2} \cr
\end{pmatrix}
=
\begin{pmatrix}
F(r)Y^{j, m}_{j-1/2} \left(\theta, \phi \right) \cr
-if(r)Y^{j, m}_{j+1/2} \left(\theta, \phi \right) \cr
\end{pmatrix},
\end{equation}
$ Y^{j, m}_{j \pm 1/2} \left(\theta, \phi \right) $ are the common eigenspinor-function of $\hat{j}_{z}$ and $\hat{J}^{2}$ and
\begin{equation}
F(r) = x^{\gamma} e^{-ax} \sum_{\nu = 0}^{n^{\prime}} a_{\nu} x^{\nu},
\end{equation}
\begin{equation}
f(r) = x^{\gamma} e^{-ax} \sum_{\nu = 0}^{n^{\prime}} b_{\nu} x^{\nu},
\end{equation}
where
\begin{equation}
n^{\prime} = n - \left( j + \frac{1}{2} \right),
\end{equation}
\begin{equation}
\gamma = -1 + \sqrt{\left( j + \frac{1}{2} \right)^{2} - \alpha^{2}},
\end{equation}
\begin{equation}
a =  \sqrt{1 - \frac{E^2}{m^2 c^4}},
\end{equation}
and
\begin{equation}
x = \left( \frac{mc}{\hbar} \right) r,
\end{equation}
we employ the following identity  \cite{Merzbacher},
\begin{equation}
\label{ident1}
\vec{\sigma} \cdot \hat{\bf p} = \vec{\sigma} \cdot \vec{e}_{r} \left( -i\hbar \frac{\partial}{\partial r} + i\frac{\vec{\sigma} \cdot \hat{\bf L}}{r} \right),
\end{equation}
with
\begin{equation}
\label{ident2}
\vec{\sigma} \cdot \vec{e}_{r}Y^{j, m}_{j \pm 1/2}  = - Y^{j, m}_{j \pm 1/2}.
\end{equation}

After some algebra we get
\begin{equation}
\label{1fi1}
\left( \vec{\sigma} \cdot \hat{\bf p} \right) \phi_{1} = i\hbar \left[ \frac{dF}{dr} - \left(j - \frac{1}{2} \right) \frac{F}{r} \right] Y^{j, m}_{j+\frac{1}{2}} ,
\end{equation}
\begin{equation}
\label{1fi2}
\left( \vec{\sigma} \cdot \hat{\bf p} \right) \phi_{2} = \hbar \left[ \frac{df}{dr} + \left(j + \frac{3}{2} \right) \frac{f}{r} \right] Y^{j, m}_{j-\frac{1}{2}} ,
\end{equation}
and
\begin{equation}
\label{2fi1}
\left( \vec{\sigma} \cdot \hat{\bf p} \right)^{2} \phi_{1} = -\hbar^{2} \left[ \frac{d^{2}F}{dr^{2}} + 2\frac{1}{r}\frac{dF}{dr} -
\left(j - \frac{1}{2} \right)\left(j + \frac{1}{2} \right) \frac{F}{r^{2}}\right] Y^{j, m}_{j-\frac{1}{2}},
\end{equation}
\begin{equation}
\label{2fi2}
\left( \vec{\sigma} \cdot \hat{\bf p} \right)^{2} \phi_{1} = i \hbar^{2} \left[ \frac{d^{2}f}{dr^{2}} + 2\frac{1}{r}\frac{df}{dr} -
\left(j + \frac{1}{2} \right)\left(j + \frac{3}{2} \right) \frac{f}{r^{2}}\right] Y^{j, m}_{j+\frac{1}{2}},
\end{equation}
and
$$
\left( \vec{\sigma} \cdot \hat{\bf p} \right)^{3} \phi_{1} =  - i\hbar^{3} \left[ \frac{d^{3}F}{dr^{3}} -
 \left(j - \frac{5}{2} \right) \frac{1}{r}\frac{d^{2}F}{dr^{2}} \right] Y^{j, m}_{j+\frac{1}{2}} +
$$
\begin{equation}
\label{3fi1}
i\hbar^{3} \left[ \left(j + \frac{1}{2} \right)\left(j + \frac{3}{2} \right)\frac{1}{r^{2}}\frac{dF}{dr} -
 \left(j - \frac{1}{2} \right)\left(j + \frac{1}{2} \right)\left(j + \frac{3}{2} \right)\frac{F}{r^{3}} \right] Y^{j, m}_{j+\frac{1}{2}},
\end{equation}

$$
\left( \vec{\sigma} \cdot \hat{\bf p} \right)^{3} \phi_{2} =  - \hbar^{3} \left[ \frac{d^{3}f}{dr^{3}} +
 \left(j + \frac{7}{2} \right) \frac{1}{r}\frac{d^{2}f}{dr^{2}} \right] Y^{j, m}_{j-\frac{1}{2}} +
$$
\begin{equation}
\label{3fi2}
\hbar^{3} \left[ \left(j - \frac{1}{2} \right)\left(j + \frac{1}{2} \right)\frac{1}{r^{2}}\frac{df}{dr}  +
 \left(j - \frac{1}{2} \right)\left(j + \frac{1}{2} \right)\left(j + \frac{3}{2} \right)\frac{f}{r^{3}} \right] Y^{j, m}_{j-\frac{1}{2}}.
\end{equation}

At last,
$$
\langle \psi | \left( \vec{\alpha} \cdot \hat{\bf p} \right)^{3} | \psi \rangle =
 - \hbar^{3}\int \left( F^{*}\frac{d^{3}f}{dr^{3}} - f^{*} \frac{d^{3}F}{dr^{3}} \right) r^2 dr
$$
$$
- \hbar^{3} \left(j + \frac{7}{2} \right)\int F^{*} \frac{d^{2}f}{dr^{2}} r dr
- \hbar^{3} \left(j - \frac{5}{2} \right) \int f^{*} \frac{d^{2}F}{dr^{2}} r dr
$$
$$
+ \hbar^{3} \left(j - \frac{1}{2} \right)\left(j + \frac{1}{2} \right) \int F^{*} \frac{df}{dr} dr
- \hbar^{3} \left(j + \frac{1}{2} \right)\left(j + \frac{3}{2} \right) \int f^{*} \frac{dF}{dr} dr
$$
\begin{equation}
\label{3fi3}
+ \hbar^{3} \left(j - \frac{1}{2} \right)\left(j + \frac{1}{2} \right)\left(j + \frac{3}{2} \right) \int \left( F^{*}f + f^{*}F \right) \frac{dr}{r}.
\end{equation}

\subsection{Ground state energy}

\h For the ground state we have $j = \frac{1}{2}$ and $n^{\prime} = 0$, then
\begin{equation}
F_{0}(r) = a_{0} x^{\gamma} e^{-ax},
\end{equation} 
\begin{equation}
f_{0}(r) = b_{0} x^{\gamma} e^{-ax},
\end{equation}
where 
\begin{equation}
\gamma = \epsilon - 1,
\end{equation}
\begin{equation}
a_{0} = \left( \frac{2a}{b} \right)^{\gamma + 1} \sqrt{\frac{\left(1 + \epsilon \right)}{\Gamma \left( 2\gamma +3 \right) }},
\end{equation}
\begin{equation}
b_{0} = \sqrt{\frac{1 - \epsilon}{1 + \epsilon}}a_{0},
\end{equation}
with $\epsilon = \sqrt{1- \alpha^2}$.

Hence
\begin{equation}
\langle \psi_{0} | \left( \vec{\alpha} \cdot \hat{\bf p} \right)^{3} | \psi_{0} \rangle =
 m^{3}c^{3}\frac{\left( 1- \epsilon^{2} \right)^{2}}{\epsilon \left(2 \epsilon - 1 \right)},
\end{equation}
where $E_{0} = mc^2 \epsilon$ is the energy of the $| \psi_{0} \rangle $ ground state of the hydrogen atom obtained from the ordinary Dirac equation.



\end{document}